\documentstyle[12pt]{article}

\def\be{\begin{equation}}
\def\ee{\end{equation}}
\def\bea{\begin{eqnarray}}
\def\eea{\end{eqnarray}}
\def\nen{\nonumber \\ \relax}

\def\forcepar{{\hskip 10pt\vskip -15pt}}

\newfont{\headfont}{cmbx10 scaled 1440}
\newfont{\namefont}{cmr10}
\newfont{\initialfont}{cmr10 scaled 1200}
\newfont{\addfont}{cmti7 scaled 1440}
\newfont{\boldmathfont}{cmbx10}
\newfont{\figfont}{cmr7 scaled 1200}

\def\seq{\ =\ }
\def\pls{\ +\ }
\def\mi{\ -\ }

\def\half{\frac{1}{2}}
\def\inv#1{\frac{1}{#1}}

\def\cb{{\cal B}}

\def\cd{{\cal D}}

\def\ct{{\cal T}}

\def\cz{{\cal Z}}

\def\IR{{I \kern -0.4em R}}
\def\IC{{I \kern -0.65em C}}

\def\np#1{{\it Nucl. Phys.} {\bf B#1}}
\def\cmp#1{{\it Commun. Math. Phys.} {\bf #1}}
\def\ijmp#1{{\it Intl. J. Mod. Phys.} {\bf A#1}}

\def\mpl#1{{\it Mod. Phys. Lett.} {\bf A#1}}

\def\pl#1{{\it Phys. Lett.} {\bf #1B}}

\newfont{\headfontb}{cmbx10 scaled 1728}
\def\spbf{{super-$BF$}}
\def\sbf{{SUSY-$BF$}}

\def\cj{{\cal J}}
\def\cq{{\cal Q}}
\def\dual{^\star\!}

\def\Xhat{\bf{\widehat X}}
\def\Yhat{\bf{\widehat Y}}

\def\trans#1#2#3{[{#1},{#2}\}\seq{#3}}
\def\sigharm{\leavevmode\raise .72em\hbox{\the\scriptfont0 $\circ$}
            \kern-.59em\hbox{\the\scriptfont0 $\Sigma$}}

\def\de{\nabla}
\def\sp#1{{}^{#1}}                              
\def\sb#1{{}_{#1}}                              
\def\a{\alpha}
\def\b{\beta}

\def\d{\delta}
\def\e{\epsilon}

\def\g{\gamma}

\def\s{\sigma}

\def\fracm#1#2{\hbox{\large{${\frac{{#1}}{{#2}}}$}}}

\begin{document}
\begin{titlepage}
\renewcommand{\thefootnote}{\fnsymbol{footnote}}
\begin{center}
{\headfontb Extended Supersymmetry\\ and  Super-BF Gauge
Theories}\footnote{This work is supported in part by funds
provided by the
U. S. Department of Energy (D.O.E.) under cooperative agreement
\#DE-FC02-94ER40818.}

\end{center}
\vskip 0.3truein
\begin{center}
{
{\Large R}{OGER} {\Large B}{ROOKS}
{ AND}
{\Large S.} {\Large J}{AMES} {\Large G}{ATES},
{\Large J}{R.}\footnote{Permanent address:
Department of Physics,
University of Maryland,
College Park, MD 20742-4111  USA.  Research supported by NSF grant
NSF-PHY-91-19746}
}
\end{center}
\begin{center}
{\addfont{Center for Theoretical Physics,}}\\
{\addfont{Laboratory for Nuclear Science}}\\
{\addfont{and Department of Physics,}}\\
{\addfont{Massachusetts Institute of Technology}}\\
{\addfont{Cambridge, Massachusetts 02139 U.S.A.}}
\end{center}
\vskip 0.5truein
\begin{abstract}
The absence of fermion kinetic terms in supersymmetric-$BF$ gauge theories
is established.  We do this by means of explicit off-shell (superspace)
constructions.  As part of our study we give the
superspace constraints for D=3, N=4 super Yang-Mills along with the D=3,
N=4 superconformal algebra.   The puzzle we are interested in solving is
the fact that the topological cousins, known as super-$BF$ gauge theories,
of certain supersymmetric-$BF$ theories have kinetic terms for the twisted
fermions.
We show that the map which takes the latter to the former includes a Hodge
decomposition of the twisted fermions.  In conjunction with this result, we
argue that it is natural to modify the naive path integral measure of
supersymmetric-$BF$ theories to include the Ray-Singer analytic
torsion.
\vskip 0.5truein
\leftline{CTP \# 2339  \hfill July 1994}
\leftline{UMDEPP-95-07}
\smallskip
\leftline{hep-th/xxxxxxx}
\end{abstract}
\vskip 0.5truein
\end{titlepage}
\setcounter{footnote}{0}
\section{Introduction}
\medskip\forcepar
In this paper, we will construct and study various applications of
extended supersymmetric $BF$ gauge theories\footnote{For a review of
non-supersymmetric $BF$ theories and topological field theories (TFTs)
in general, see reference \cite{TFT_rev}.} (SUSY-$BF$) and D=3,
N=4 supersymmetry in general.  As one of the applications, we will
show that a pair of problems associated with supersymmetric $BF$ gauge
theories (\sbf) share a common solution in the form of an insertion of
the Ray-Singer (R-S) torsion \cite{torsion} in the measure of the path
integral.

The first of these problems, as we will see, is the lack of fermionic
determinants to cancel those from the bosons in these supersymmetric
theories.  Indeed, we will show that the fermionic contribution to the
latter is only as an off-diagonal mass term.  Although our explicit
constructions will be primarily in three dimensions, we expect the
results to hold for arbitrary dimensions.

The second problem arises as follows.  It is well known that a large
class of topological quantum field theories (TQFTs) may be obtained
by twisting certain supersymmetric field theories.  However, this
procedure does not work for SUSY-$BF$ theories.  The supposed twisted
cousins of these theories, which we shall call super-$BF$
theories\footnote{In order to avoid confusion in nomenclature we will
refer to the ordinary supersymmetric $BF$ theories as \sbf{} while the
corresponding TQFTs will be called super-$BF$ theories.}, have kinetic
terms for the would-be twisted fermions.  Thus far, super-$BF$ theories
have been constructed only via BRST gauge fixing.

To solve the first problem we will simply insert the Ray-Singer analytic
torsion in the measure of the path integral for the SUSY-$BF$ gauge theory.
What is more interesting, we find that such an insertion is also needed
in order to solve the second problem.  In order to have a match between
the twisted SUSY-$BF$ and super-$BF$ partition functions, one of the
fermions obtained by twisting the spin-$\half$ fields in the former
theory must be Hodge decomposed.  This change of variables results
in the addition of the R-S torsion to the measure of the path
integral\footnote{Recall that for closed manifolds, the R-S torsion
is purely topological.}.  This may be interpreted as defining a new
vacuum for the SUSY-$BF$ theory.

The paper is organized as follows.  We begin in sections \ref{N1SYM}
(for D=3, N=1 and D=4,N=1) and \ref{N4SYM} (for D=3, N=4) by
illustrating the presence of only (off-diagonal) mass terms for the
fermions in SUSY-$BF$ theories.  This lack of kinetic terms for the
fermions comes as no surprise due to two statements.  First,
supersymmetric Chern-Simons (CS) theories \cite{OurCS} are known to
have only these mass terms due to the fact that the CS action give
a mass for the gauge field.  Second, $BF$ theories with gauge group
$G$ may be obtained from CS theories \cite{Wit} via the formation of
an inhomogeneous gauge group out of $G$.  Our motivation for
constructing these theories is that we will later use them to
illustrate the problem in twisting from SUSY-$BF$ to super-$BF$ theories.
Additionally, our constructions of superspace geometries for D=3, N=4
super Yang-Mills  will not only fill in a gap in the literature but will
also serve to point out the richness of supermultiplets for the latter
theories.    Next, in section \ref{Novel}, we illustrate some novel
features of SUSY-$BF$ theories amongst which is the existence of a
``minimal" N=4 action.  Then in section \ref{Twist}, we twist the minimal
D=3, N=4 SUSY-$BF$ theory to obtain a topological gauge theory with
a mass term for Grassmann-odd one- and two-form fields.  We then show
that in order to obtain the super-$BF$ theory we must Hodge decompose
the two-form field.  Our conclusions may be found in section \ref{Conc}.
As part of our general formulation of D=3, N=4 supersymmetric theories,
we discuss the construction of off-shell D=3, N=4 supergravity  and give
the D=3, N=4 superconformal algebra in appendices A and B, respectively.
Then, commensurate with our discussion of SUSY-$BF$ theories, a
proposed a new action for N=4 U(1) supersymmetric anyons is given in
appendix C.
\vskip 0.5truein
\section{Generic Properties of  SUSY-$BF$}\label{N1SYM}
\setcounter{equation}{0}
\medskip
\forcepar
In order to set the stage for our discussions let us begin by
constructing, as exercises, the SUSY-$BF$ N=1 supersymmetric actions
in three and four space-time dimensions\footnote{The notation $S_{SUSY}
[D,N]$ is used to denote the (untwisted) $D$-dimensional, $N$-extended
supersymmetric $BF$ action.}.
\bigskip
\subsection{D=3}
\medskip\forcepar
N=1 superspace construction of the SUSY-$BF$ action proceeds as follows.
First, introduce a spin-$\half$ superfield, $\cb_\alpha$, along with the
spinor field-strength, $W_\alpha$,  of the super Yang-Mills theory
\cite{Superspace,OurCS} and write the superfield action
\be
S_{SUSY}[3,1]\seq -\half \int d^3\sigma d^2\theta\, Tr(\cb^\alpha
W_\alpha)\
\ .\ee
Then with the component fields given by projection as
\bea
\cb_\alpha|&\seq& \beta_\alpha\ \ ,\qquad \nabla_\beta \cb_\alpha|\seq
i(\gamma^a)_{\alpha\beta} B_a \pls C_{\beta\alpha}b\ \ ,\quad \nabla^2
\cb_\alpha|\seq \rho_\alpha\ \ ,\nen
W_\alpha|&\seq& \psi_\alpha\ \ ,\qquad \nabla_\beta W_\alpha|\seq i\half
\epsilon^{abc} (\gamma_a)_{\alpha\beta} F_{bc}\ \ ,\nen
 \nabla^2 W_\alpha|&\seq& -i (\gamma^a)_{\alpha\beta} \cd_a \psi^\beta\ \
,\eea
where $\cd_a$ is the Yang-Mills covariant derivative,
we find the action to be
\be
S_{SUSY}[3,1]\seq \half \int d^3\sigma\, Tr (\epsilon^{abc} B_a
F_{bc} \pls
i\beta^\alpha (\gamma^a)_{\alpha\beta} \cd_a\psi^\beta\mi \rho^\alpha
\psi_\alpha)\ \ .\ee
The Dirac action in this expression is actually fictitious.  This is seen
as follows.  Recall that the spinor field-strength of D=3, N=1 super
Yang-Mills satisfies the Bianchi Identity, $\nabla^\alpha W_\alpha=0$.
This implies that the superspace action is invariant under the superfield
transformation $\delta \cb_\alpha =\nabla_\alpha \Lambda$.  Writing,
$\Lambda|=X$, $\nabla_\alpha \Lambda|=X_\alpha$ and $\nabla^2\Lambda|=X'$,
we see that the component transformations include $\delta \beta_\alpha=
X_\alpha$, $\delta B_a=\cd_a X$ and $\delta b=-X'$.  Consequently,
$\beta_\alpha$ and $b$ may be set to zero algebraically.  Hence, in
this Wess-Zumino gauge, the component action reads
\be
S_{SUSY}[3,1]\seq \half \int d^3\sigma\, Tr (\epsilon^{abc} B_a
F_{bc}\mi
\rho^\alpha \psi_\alpha)\ \ .\ee
Observe that the fermions only appear in an off-diagonal mass term.

There is another feature of SUSY-$BF$ theories which we would like to
point out.  Our (real) N=1 supersymmetric action is actually invariant
under a complex N=1 supersymmetry transformation.  This is best seen
in superfield form for which the action is symmetric under the
interchange of  $\cb_\alpha$ and $W_\alpha$.  Of course, these two
superfields form representations of two completely different
supermultiplets, thus we do not expect this symmetry to be
preserved (off-shell) at the level of the algebra.  Nevertheless,
we find that the component action is invariant under the complex
supersymmetry transformations generated by $Q_\alpha$ and $\bar
Q_\alpha$ for which
\bea
[Q_\alpha,A_a]&\seq& (\gamma_a)_\alpha{}^\beta(\psi_\beta\pls
i\rho_\beta)\ \
,\nen
\{Q_\alpha,\psi_\beta\}&\seq& -\epsilon^{abc}(\gamma_a)_{\alpha\beta}
(F_{bc}\mi i\cd_{[b}B_{c]})\ \ ,\nen
[Q_\alpha,B_b]&\seq&i(\gamma_a)_\alpha{}^\beta(\psi_\beta\mi
i\rho_\beta)\ \
,\nen
\{Q_\alpha,\rho_\beta\}&\seq&-i\epsilon^{abc}(\gamma_a)_{\alpha\beta}(F_
{bc}\pls i \cd_{[b}B_{c]})\ \ .\eea
Now this complex N=1 supersymmetry does not form a N=2 supersymmetry
algebra
off-shell.  A quick check of this statement follows from computing
$\{Q_\alpha,Q_\beta\}$ acting on $B_a$ and finding that it is proportional
to $F_{ab}$, the field strength of $A_a$.  As $F_{ab}=0$  is the equation
with follows from varying the action with respect to $B_a$, we see that
this anti-commutator vanishes only on-shell.  In order to elevate this
on-shell supersymmetry representation  to that of an $N=2$ off-shell
supersymmetry we must include additional auxiliary fields.  In the next
section we will skip off-shell N=2 supersymmetry and move directly to
off-shell N=4 supersymmetry which will be of interest to us not for the
purposes of manifesting the above properties (even though they will also
be apparent there) but because of its expected relation to super-$BF$
theories.  The fact that, classically, supersymmetry requires the fermion
to vanish is related to the (non-)existence of reducible flat connections
as follows.  By taking a supersymmetry transformation of the equation of
motion for the connection, $F_{ab}=0$, we find that this implies that the
spinors must be gauge-covariantly constant\footnote{A similar result holds
for the $B_a$ field and its super-partner.}:
\be
0\seq[Q_\alpha, F_{ab}]~\longrightarrow~\cd_a \psi_\alpha\seq 0\ \ ,\ee
after making  use of the properties of the D=3 gamma-matrices.  Since
this condition holds for any $\alpha$, we see that if any solutions
of the equation $\cd_a\Phi=0$ existed (where $\Phi$ is some Lorentz
scalar which transformations in the adjoint representation of the gauge
group), then there would be  non-zero solutions for $\psi_\alpha$.
Thus far we have only used the equation of motion for the gauge field
and its supersymmetry transformation.  Now, the equation of motion for
the fermions impose that they vanish.  Thus the existence of reducible
connections would at least lead to novel classical behavior for the
SUSY-$BF$ theory.  This argument, assumes that the Lorentz spin-connection
is zero.  It would be interesting to extrapolate it to the case of
arbitrary curved manifolds with attention focused on  the existence
of global supersymmetries.
\bigskip
\subsection{D=4}
\medskip\forcepar
It appears that the behaviour we saw above is a universal feature of
SUSY-$BF$ theories.  As further evidence of this we now turn to the four
dimensional theory.

In analogy with three dimensions, in four dimensions we introduce a chiral
spinor superfield, $\cb_\alpha$ along with the spinor superfield-strength,
$W_\alpha$.  The action is roughly of the same form as that of
$S_{SUSY}[4,1]$:
\be
S_{SUSY}[4,1]  \seq -\half \int d^4\sigma d^2\theta\, Tr(\cb^\alpha
W_\alpha)
\pls (\rm c.c.)\ \ .\ee

Since $W_\alpha$ satisfies the Bianchi Identities $\nabla^\alpha W_\alpha+
{\bar \nabla}^{\dot\alpha}{\bar W}_{\dot\alpha}=0$ and
${\bar\nabla}_{\dot\alpha}W_\alpha=0$, we  find the action to be invariant
under separate transformations for $B_\alpha$:
\be
\delta B_\alpha \seq \nabla_\alpha \Lambda\ \ ,\qquad
\delta B_\alpha\seq
(\sigma^a)_{\alpha}{}^{\dot\beta}{\bar\nabla}_{\dot\beta}\Lambda_a\ \
.\label{EqnN4Sym}\ee
In these expressions, $\Lambda$ is a chiral, scalar superfield parameter and
$\Lambda_a$ is an anti-chiral superfield: $\nabla_\alpha\Lambda_a=0$.

Now, the components of $B_\alpha$ are of the same form as before with the
exception that
\be
\nabla_\alpha B_\beta\seq i(\sigma^{ab})_{\alpha\beta} B_{ab}\pls
C_{\alpha\beta} b\ \ .\ee
Then, much as in the three dimensional case, we see from the first
transformation in eqn. \ref{EqnN4Sym}, that we can set $\beta_\alpha=0$.
However, unlike three dimensions, not all of $b$ can be set equal to
zero. (The $B_\alpha$ multiplet is actually the dilaton multiplet
of N = 1 four dmensional string theory. The part of $b$ that
we cannot set to zero corresponds to the would-be dilaton.)
The two form symmetry: $\delta B_{ab}=\cd_{[a}\xi_{b]}$ follows
from the second superfield transformation.

The component lagrangian differs from that of three dimensions in only
two respects.  First, the field $B$ is now a two form (with attendant
D=4 Levi-Cevita tensor) and the fermion term is now of the
form $\rho^\alpha\psi_\alpha+(c.c)$.  Hence, as before, we expect that
this action is actually invariant under a pair of N=1 supersymmetries.
\bigskip
\subsection{Path Integral Measure}
\setcounter{equation}{0}
\medskip
\forcepar
Thus far, our constructions have been purely classical.  We have obtained
\sbf actions without derivative terms for fermions.  As was discussed
above, this means that the classical supersymmetry theory is the same
(modulo the presence of reducible connections) as the non-supersymmetric
theory.  It also appears that the quantum theories will be the same.
However, this goes against our prior experience with supersymmetric
theories.  Up to signs, the bosonic determinants are supposed to
be cancelled by those from fermions.   That does not occur here.  Thus,
we propose to modify the measure of the path integral of \sbf so that
this feature of supersymmetric theories will be maintained.  In this case,
it is easily accomplished as follows. Since the ratio of the non-zero mode
determinants which arise \cite{TFT_rev} from the integration over the
connection and $B$ fields is equal to that which appears in the inverse
R-S torsion  valued in the flat connections, $A_0$, we simply insert the
R-S torsion, $T[A_0]$ as part of the definition of the flat connection
part of the measure for \sbf gauge theories: $[dA_0]\to [dA_0]T[A_0]$.
Since the partition function has support only on flat connections, $A_0$
in $T[A_0]$ may  -- in turn --be replaced by the general connection, $A$.
Hence we propose that the path integral for N=1 \sbf theories is
\be
{\cal Z}_{SUSY} \seq \int[dA] [dB][d\rho][d\psi] T[A]
e^{-S_{SUSY}[D,1][A,B,\rho,\psi]}\ \
,\label{ZSBF}\ee
where
$$T[A]\seq \det{}^{-3/2}\triangle_A^{(0)}\det{}^{1/2}\triangle_A^{(1)}\ \
,$$
regardless of the dimension of space-time. Here, $\triangle_A^{(k)}$ is
the covariant laplacian on $k$-forms.  Since $T[A_0]$ is topological it
does not introduce any local degrees of freedom into the theory.

Having made this insertion, we must determine whether or not the
supersymmetries are still preserved in eqn. (\ref{ZSBF}).  To answer this
question we first recall \cite{TFT_rev} that $T[A]$ may we written as the
path integral over a certain set of fields of the exponential of the
Gaussian action for these fields in the gauge background given by the
connection, $A$.  Then the coupling of $A$ to these fields in the modified
action is of the form $Tr(A\cdot J)$; where $J$ is the gauge current for
the additional fields.  Consequently, we can maintain the supersymmetries
by declaring that the new fields, hence $J$, do not transform under the
latter.  This means that the supersymmetry transformation of the action
which is used to represent the R-S torsion is of the form $Tr(\delta A
\cdot J)$.  Such a term may be cancelled by re-defining the supersymmetry
transformation law of the  fermion which does not appear in $\delta A$ to
be proportional to the current.  To conclude this section, we summarize
it and offer the following road map for the remainder of the paper.  In
this section, we have seen that the off-shell N=1 supersymmetric BF actions
are actually invariant under a pair of these supersymmetries.  However,
these  transformations do not form a N=2 supersymmetry algebra, off-shell.
Next, we will turn to the formulation of D=3, N=4 super Yang-Mills.  Then,
in section 4, we will demonstrate that for SUSY-$BF$ theories, N=4
supersymmetric actions can be constructed using ``reduced" supermultiplets.
This is due to the presence of additional local symmetries in these
theories which do no have derivative terms for fermions.  We then find
that it is these theories which can be twisted (via a Hodge decomposition
procedure) to super-$BF$ theories.
\vskip 0.5truein
\section{D=3, N=4 Super Yang-Mills}\label{N4SYM}
\setcounter{equation}{0}
\medskip
\forcepar
There are two off-shell 3D, N = 4 vector supermultiplets.  This is
an example of a phenomenon that was noted a long time ago in
supersymmetric theories \cite{GT1}.  Namely it is often the case that
for a given set of propagating fields (on-shell theory), there is one
or more distinct off-shell theories.  These different off-shell
representations of the same physical states are called ``variant
representation'' of the supermultiplet.  Sometimes, but not always,
variant representations are related by a duality transformation.

The most powerfully known consequence of the existence of variant
representations is the occurrence of ``mirror symmetry'' in compactified
heterotic string theory \cite{Dix}. One way to view mirror symmetry is
that it maps a particular off-shell supermultiplet into one
of its variant representation and vice-versa.  In two dimensional
N = 2 theories, this is the situation obtained in heterotic string
theories that contain the chiral scalar multiplet and its variant
(the twisted chiral multiplet)\cite {GaHuR}. The fact that we have
discovered the existence of a previously unsuspected 3D, N = 4 vector
supermultiplet suggest the exciting possibility of extending the concept of
mirror symmetry to three dimensions!

Let us start with a 4D, N = 2 vector supermultiplet and a 4D, N = 2 tensor
supermultiplet.  The off-shell representations of both of these theories
have been known for a long time. Working with the usual actions in four
dimensions shows that these two supermultiplets both describe the propagation
of 4 bosonic degrees of freedom and 4 fermionic degrees of freedom.  Now
consider a toroidal compactification to three dimensions.  This will
necessarily split off some of the components of the gauge fields in each
supermultiplet into different 3D representations of the SO(1,2) group.
Under the dimensional reduction,  the 4D vector gauge field ``yields'' a
3D vector gauge field as well as one scalar.  This process has been called
``the inverse Higgs phenomenon.'' Similarly, the 4D antisymmetric tensor
gauge field ``yields'' a 3D, 2-form gauge field as well as one 3D vector
gauge field.  In three dimensions, a 3D 2-form gauge field propagates no
physical degrees of freedom.  By a duality transformation, it can be replaced
by an auxiliary scalar.  Also, the number  of supersymmetries for the models
double because an irreducible spinor representation of SO(1,3) contains two
irreducible spinor representations  of SO(1,2). So the two models that result
are 3D, N = 4 vector supermultiplet models.  Thus, we arrive at the possible
existence of two distinct 3D, N = 4 vector supermultiplets!

One of the two distinct 3D, N = 4 off-shell vector supermultiplets is given
by the following commutator algebra.
$$
[\de \sb{\a i} , \de \sb{\b j} \}  = i 4 g C_{\a \b} C_{ i j }
{\bar {~W~}}^{\cal I} t_{\cal I} ~~,
{}~~~~~~~~~~~~~~~~~~~~~~~~~~~~~~~$$
$$
{}~~~~~~~~~~~~~~~~~~~~ [\de \sb{\a i} , \bar \de \sb{\b} {}^j \} =  i 2 \d_i
{}^j
(\g \sp{c})  \sb{\a \b} \de \sb{c} ~+~ 2 g \d_i {}^j C_{\a \b}  S^{\cal I}
t_{\cal I} ~~, ~~~~~~~~~~~~~~~~~~~~~~~~~~~~~~~~~~~ $$
$$
[\de \sb{\a i} , \de \sb{b} \} = g (\g_b)_\a {}^\d
{\bar W}_{\d i} {}^{\cal I} t_{\cal I} ~~,
{}~~~~~~~~~~~~~~~~~~~~~~~~~~~~~~~~~~~$$
\begin{equation}
[\de \sb{a} , \de \sb{b} \} = i g F_{a b} {}^{\cal I} t_{\cal I}
{}~~~.~~~~~~~~~
{}~~~~~~~~~~~~~~~~~~~~~~~~~~~~~~~~
\end{equation}
The following equations must be imposed in order to satisfy the Bianchi
identities,
$$~~~~~~\de_{\a i } {\bar {~W~}}^{\cal I} ~=~ 0 ~~~,
{}~~~~~~~~~~~~~~~~~~~~~~
{}~~~~~~~~~~~~~~~~~~~~~~~~~~~~~~~~~~~~~~~~~~~~~~~~~~~~~~~$$
$$~~~~~~{\bar \de}_{\a}{}^{ i } {\bar {~W~}}^{\cal I} ~=~  C^{ i j }
{\bar W}_{\a j} {}^{\cal I} ~~~,
{}~~~~~~~~~~~~~~~~~~~~~~~~~~~~~~~~~~~~~~~~
{}~~~~~~~~~~~~~~~~~~~~~~~~~~~$$
$${ \de}_{\a i } {S}^{\cal I} ~=~  - i {\bar W}_{\a i} {}^{\cal I} ~~~,
{}~~~~~~~~~~~~~~~~~~~~~~~~~~~~~~~~~~~~~~~~~~~~~~~~~~~~~~~~
{}~~~$$
$$~~~~~~~~~\de_{\a i } {\bar W}_{\b j} {}^{\cal I} ~=~ i 2  C_{ i j } (\g
\sp{c})  \sb{\a \b} (\de_c {\bar {~W~}}^{\cal I} ) ~+~ 2 g  C_{ i j } C_{\a
\b}
\left [ ~ S , {\bar {~W~}} \right ]^{\cal I} ~~,
{}~~~~~~~~~~~~~~~~~~~~~~~~$$
$${\bar \de}_{\a}{}^{ i } {\bar W}_{\b j} {}^{\cal I} ~=~ i  \d_{j}{}^i (\g
\sb{a})  \sb{\a \b} \left [\fracm 12 \e^{a b c } F_{b c} {}^{\cal I} (A)
{}~+~ i  ( \de^a  S^{\cal I} ) \right ] ~+~ i  C_{\a \b} d^{\cal I}
\sb j {}^i  ~~, ~~~~~~~~ $$
$$
\de_{\a i } d^{\cal I} \sb j {}^k ~=~ - [  2 \d_i \sp k \d \sb j \sp l -
\d_j \sp k \d \sb i \sp l ] [ ~(\g \sp{c})  \sb{\a} \sp{\b}
 (\de_c {\bar {~W~}_{ \b l} }^{\cal I} ) ~+~ g \left [  {\bar {~W~}_{ \a l} } ,
 S \right ] {}^{\cal I} ~~~~~ $$
\begin{equation}
 ~~~~~~~~~~~~~~~~~~~~~~~~~~+~g  C_{l r } \left [ W_{\a}{}^{r} ,{\bar
{~W~}} \right ] {}^{\cal I} ~] ~~~~.
\end{equation}
The 3D vector multiplet, we have just discussed is the one that
``descends'' from the 4D, N = 2 vector supermultiplet.  The component
level equations of motion that follow from the usual action for
a vector gauge field begin by setting $ d^{\cal I} \sb j {}^k = 0$.
The first spinorial derivative of this restriction implies the
Dirac-like equation for the spinor and the second spinorial derivatives
imply the Yang-Mills and Klein-Gordon equations for the bosons.

The second 3D, N = 4 vector supermultiplet has a covariant derivative
whose commutator algebra takes the form,
$$
[\de \sb{\a i} , \de \sb{\b j} \}  = 0 ~~,
{}~~~~~~~~~~~~~~~~~~~~~~~~~~~~~~~
{}~~~~~~~~~~~~~~~~~~~$$
$$
{}~~~~~~~~~~~~~~~~~~~~ [\de \sb{\a i} , \bar \de \sb{\b} {}^j \} =  i 2 \d_i
{}^j
(\g \sp{c})  \sb{\a \b} \de \sb{c} ~+~ 2 g  C_{\a \b}  \varphi^{\cal I}
{}_i {}^j t_{\cal I} ~~, ~~~~~~~~~~~~~~~~~~~~~~~~~~~~~~~~~~~ $$
$$
[\de \sb{\a i} , \de \sb{b} \} = g (\g_b)_\a {}^\d {\bar U}_{\d i} {}^{\cal I}
t_{\cal I} ~~, ~~~~~~~~~~~~~~~~~~~~~~~~~~~~~~~~~~~$$
\begin{equation}
[\de \sb{a} , \de \sb{b} \} = i g F_{a b} {}^{\cal I} t_{\cal I}
{}~~~.~~~~~~~~~
{}~~~~~~~~~~~~~~~~~~~~~~~~~~~~~~~~
\end{equation}
The following equations must be imposed in order to satisfy the Bianchi
identities,
$$~~~~~~~~~~~~~~~~ \varphi^{\cal I} {}_i {}^i  ~=~ 0 ~~~, ~~~
\varphi^{\cal I}
{}_i {}^j ~-~ ({ { \varphi^{\cal I} {}_j {}^i }})^* ~=~ 0 ~~~,
{}~~~~~~~~~~~~
{}~~~~~~~~~~~~~~~~~~~~~~~~~~~~~~~~~~~~~~~~~~~~~~
{}~~~$$
$$~~~~~~~~{\de}_{\a i } \varphi^{\cal I} {}_j {}^k ~=~  - i \left [
 2 \d_i {}^k {\bar U}_{\a j}{}^{\cal I} ~-~ \d_j {}^k {\bar U}_{\a i}
{}^{\cal I} \right ] ~~~,
{}~~~~~~~~~~~~~~~~~~~~~~~~~~~~~~~~~~~~~~~~
{}~~~~~~~~~~~~~~~~~$$
$$~~~~~{ \de}_{\a i } {\bar U}_{\b j} {}^{\cal I} ~=~ -  2 C_{\a \b} C_{ i j}
{\bar J}^{\cal I}~~~, ~~~~~~~
{}~~~~~~~~~~~~~~~~~~~~~~~~~~~~~~~~~~~~~~~~~~~~~~~~~~~~~~~~
{}~~~$$
$$ ~~~~~~{ \de}_{\a i } {\bar J}^{\cal I} ~=~ 0 ~~~,~~~ a^{\cal I} ~-~
({a}^{\cal I})^* ~=~ 0 ~~~, ~~~~~~~~~~~~~~~~~~~~~~~~~~~~~
{}~~~~~~~~~~~~~~~~~~~~~$$
$$~{\bar \de}_{\a}{}^{ i } {\bar U}_{\b j} {}^{\cal I} ~=~ i
(\g \sb{a})  \sb{\a \b} \left [\fracm 12 \d_{j}{}^i  \e^{a b c } F_{b c}
{}^{\cal I} (B) ~+~ i  ( \de^a  \varphi^{\cal I} {}_j {}^i ) \right ]
{}~~~~~~~~~~~~~~~~~~~~~~~~~~~~~~~~~ $$
$$
 ~-~ i C_{\a \b} [~ a^{\cal I} \d \sb j {}^i -
\fracm 12 g \left [ \varphi {}_j {}^k , \varphi {}_k {}^i \right ]
{}^{\cal I}  ~]  ~~~. ~~~~~~~~~~~~~~~~~~~~~
$$
$$
{}~~ {\bar \de}_{\a}{}^{ i } {\bar J}^{\cal I} ~=~  i C^{ i j}
\left \{  (\g^a)_\a {}^\b ( \de_a  {\bar U}_{\b j} {}^{\cal I} ) ~-~ i g
\left [ \varphi {}_j {}^k ,  {\bar U}_{\a k } \right ]
{}^{\cal I} \right \} ~~~,~~~~~~~~~~~~~~~~~~~~~~~~~
$$
\begin{equation}
{}~~~~~~~~{ \de}_{\a i } a^{\cal I}  ~=~   \left \{ (\g^a)_\a {}^\b ( \de_a
{\bar U}_{\b i}
{}^{\cal I} ) ~-~ i g
\left [ \varphi {}_i {}^j , {\bar U}_{\a j }  \right ] {}^{\cal I}
\right \} ~~~.~~~~~~~~~~~~~~~~~~~~~~~~~~~~~~~~~~~
\end{equation}
This 3D, N = 4 vector multiplet is the one that ``descends'' from the 4D,
N = 2, 2-form supermultiplet. The component level equations of motion that
follow from the usual action for a vector gauge field begin by setting
$ a^{\cal I} = {\bar J}^{\cal I} = 0$.  The first spinorial derivative
of these restrictions imply the Dirac-like equation for the spinor and
the second spinorial derivatives imply the Yang-Mills and Klein-Gordon
equations for the bosons.

We should mention that it is actually only the Abelian version of this
theory that is obtained by dimensional reduction.  The process of replacing
the 3D, 2-form by a scalar using a duality transformation is purely a 3D
concept. It is crucial to do this for the existence of the non-Abelian version
of this theory.  It is a highly nontrivial check on the consistency of this
second unexpected 3D, N = 4 vector supermultiplet that the commutator algebra
closes without equations of motion.  We have explicitly verified this for
the spin-0 and spin-1/2 fields.  In this, the following identities are useful
$$ \e^{a b c } (\g_b)_{\a \b} (\g_c)_\g {}^\d ~=~ - i C_{\a \g} (\g^a)_\b {}^\d
- i (\g^a)_{\a \g} \d_{\b} {}^\d ~~~,~~~~~~~~~~~~~~~~~~~~~~~~ $$
$$ \left [ \varphi {}_i {}^j , \varphi {}_k {}^l \right ] ~=~
\fracm 12 \d {}_k {}^j \left [ \varphi {}_i {}^r , \varphi {}_r {}^l \right ] -
\fracm 12 \d {}_i {}^l \left [ \varphi {}_k {}^r , \varphi {}_r {}^j \right ]
{}~~~,~~~~~~~~~$$
\begin{equation}
{}~~\left [ \varphi {}_i {}^j , {\bar U}_{\a k }  \right ] ~=~
\left [ \varphi {}_k {}^j , {\bar U}_{\a i }  \right ] + \d {}_k {}^j
\left [ \varphi {}_i {}^l , {\bar U}_{\a l }  \right ] - \d {}_i {}^j
\left [ \varphi {}_k {}^l , {\bar U}_{\a l }  \right ] ~~~.
\end{equation}

It turns out that these two distinct supermultiplets are dual to each other
in a sense.  Notice that the following sets of equations are valid for
the physical fields of the first vector supermultiplet and the auxiliary
fields of the second vector supermultiplet,
$${ \de}_{\a i } {\bar W}^{\cal I} ~=~ 0 ~~~,~
{}~~ {\bar \de}_{\a}{}^{ i } {\bar W}^{\cal I} ~=~  i C^{ i j}
{ \de}_{\a j } S^{\cal I} ~~~~~,
$$
\begin{equation}
{}~{ \de}_{\a i } {\bar J}^{\cal I} ~=~ 0 ~~~,~
{}~~~~ {\bar \de}_{\a}{}^{ i } {\bar J}^{\cal I} ~=~  i C^{ i j}
{ \de}_{\a j } a^{\cal I} ~~~~~.
\end{equation}
They ``inherit'' this duality from their 4D, N = 2 vector and 2-form
ancestors.  This sense of duality is a supersymmetric generalization of
Hodge duality that relates a p-form in D dimensions to a (D - p - 1) form
and plays a role of utmost importance in formulating an off-shell
supersymmetrically consistent 3D, N = 4 BF-theory. An explicit calculation
shows that the action given by
$$
{\cal S}_{BF}^{N = 4} ~=~ \int d^3 \s [ \fracm 12 \e^{\a b c}
B_a F_{b c} (A) ~+~ ( W^{\a i} {\bar U}_{\a i} + U^{\a i} {\bar W}_{\a i} )
{}~~~~~~~~$$
\begin{equation}
{}~~~~~~~~~~~~- S a - ( W {\bar J} +  J {\bar W} ) - \fracm 12 \varphi_i
{}^j d_j {}^i ] ~~,\label{SN4BF}
\end{equation}
(where these are component fields) is invariant under the supersymmetry
transformation laws implied by the solution to the Bianchi identities
given above for each multiplet.

{}~~~~Earlier, an investigation was undertaken \cite{NiGa} in order to study
the role of supersymmetry in Chern-Simons theories.  In this previous work,
there appeared to be a barrier to finding an off-shell 3D, N = 4 Chern-Simons
theory. The action just presented above has N = 4 supersymmetry, but
there also appear two gauge fields in the action.  One of the nice features
of an off-shell action is that it can be coupled to other superfields. In
particular, the gauge fields in the action above may be coupled to 3D,
N = 4 off-shell matter scalar supermultiplets\footnote{See our final
appendix for a discussion of such matter supermultiplets.}.  On the other hand,
the 3D, N = 4 CS action\footnote{The action given is actually a mixed theory
with two CS actions and one BF action.} presented in \cite{NiGa} was an
on-shell action! In otherwords it is not possible to couple it to 3D, N = 4
matter supermultiplets.  There is a very close relation between BF theories
and CS theories.  In fact, the reason that the BF action above exists
as a consistent off-shell theory is because the two different vector
supermultiplets used are dual to each other.  This should come as no
surprise.  Afterall, in 4D ordinary vector gauge fields and 2-form
gauge fields are dual to each other.  The 4D, N = 2 vector and 2-form
supermultiplets share this property.  Looked at in this way, the solution
to the problem of finding a 3D, N = 4 CS action is obvious.  One must
find a 3D, N = 4 vector supermultiplet that is self-dual!

{}~~~~Carrying out this search for an off-shell self-dual 3D, N = 4
theory turns out to be difficult.  To date we have found no solution.
So in the following, we briefly report the status of this problem. It
is useful to look back at how the two previous 3D, N = 4 vector
supermultiplets differ. Comparing the two different, commutator algebras,
one is first struck by the fact that the two spin-0 degrees of freedom
represented by ${\bar W} $ in the commutator algebra have ``moved.''
They no longer appear in the commutator algebra of
$\left [ \de_{\a i} , \de_{\b j} \right ]$. Instead they re-appear
in the commutator algebra of
$\left [ \de_{\a i} , {\bar \de}_{\b} {}^{j} \right ]$ in the second
version of the supermultiplet.  Now what do we mean by the first two
vector supermultiplets are ``dual'' to each other?  Well, if one looks
at the action, one notices that the ``physical fields'' of one
supermultiplet are in the same SU(2) representation as the ``auxiliary
fields'' of the other suppermultiplet and vice-versa.  So a ``self-dual''
theory would be one in which the physical fields and the auxiliary fields
both occur in the same SU(2) representation. There are two ways of doing
this. Either the physical fields and auxiliary fields are both SU(2) complex
singlets or both SU(2) triplets. The calculations performed so far suggest
that in either case this leads to the necessity of adding further auxiliary
fields.

In the next section, we will show that the off-shell supersymmetric
3D, N = 4 BF actions has some very interesting properties in regard to
topological field theory!

\vskip 0.5truein
\section{Reduced Supermultiplets and Twisting}\label{Novel}
\setcounter{equation}{0}
\medskip\forcepar
It is generally believed \cite{TFT_rev} that large classes of TQFT's may be
obtained by twisting (re-defining the Lorentz representations or generators)
extended supersymmetric field theories.  In two and four dimensions, N=2
supersymmetric theories may be twisted \cite{Wit(TYM),BroKas,EgY} to
yield TQFT's in those dimensions.  In three dimensions, N=4 supersymmetric
theories are required \cite{BDL}.  Now, \spbf gauge theories are examples
of TQFT's and they may be constructed via BRST gauge fixing \cite{TFT_rev}.
Thus we know of their existence and form.  We now ask how to obtain them
via twisting.  In this section we will focus on three dimensions.  However,
we expect that our results are equally applicable in any dimension.

The natural starting point for twisting are the supermultiplets we
constructed in the previous section.  However, such an attempt immediately
leads to  two problems.  First, we see that the action (\ref{SN4BF}) only
contains mass terms for the superpartners, whereas it is known that \spbf
has derivative terms for these fields.  Secondly, under twisting, the last
(bosonic) term in $S_{BF}^{N=4}$ leads to a mass term for a pair of vector
fields which do not exist in the \spbf theory.  Thus, it seems that we need
another action from which to start the twisting procedure.  In this section,
we will solve the second of these problems. The first will be solved in the
next section. Normally, as was done in the previous section, the construction
of supermultiplets begins by matching the  bosonic and fermionic degrees of
freedom ($dof$)  without reference to an action.  This procedure can be
misleading as we now point out.  First, the number of bosonic $dof$ which
appear in the $BF$ action is four, $dof[B]=2$, $dof[A]=2$, after accounting
for gauge symmetries.  Suppose we tried to supersymmetrize this system by
introducing   a complex, $SU(2)$ doublet, spin-$\half$ field.  Since the
number of fermionic $dof$ of this field is eight, it would appear that we
must also add an additional four bosonic degrees of freedom.  However, we
should express greater care.  The $BF$ action naively had six $dof$ which
we reduced to four due to its symmetries. Thus we learn that it is important
to ascertain the symmetries of the action for the fermionic fields we have
introduced.  From the results of the previous section, we have seen that
their action is that of a mass term without any derivatives.  Hence, their
action will be invariant under both {\it local} $SU(2)$ rotations and
{\it local} $U(1)$ transformations.  The total number of gauge parameters
included in these is four.  Thus, for a complex, $SU(2)$ doublet,
spin-$\half$ field, $\Upsilon_{\alpha i}$, whose action is a mass term,
the number of degrees of freedom is four, matching the bosonic content of
the $BF$ action.  Thus, we write the action:
\be
S_{BF,red}^{N=4}\seq \int d^3\sigma\, Tr[\half \epsilon^{abc} B_a
F_{bc}\pls
\Upsilon^{\alpha i}\bar \Upsilon_{\alpha i}]\ \ . \label{SN4red}\ee
It is invariant under the set of rigid transformations
\bea
[Q_{\alpha i},A_a]&\seq& (\gamma_a)_\alpha{}^\beta \Upsilon_{\beta i}\ \
,\qquad \{Q_{\alpha i},\bar\Upsilon_{\beta j}\}\seq
(\gamma_a)_{\alpha\beta}
\epsilon^{abc}\cd_b B_a C_{ij}\  \ ,\nen
[\bar Q_{\alpha i},B_a]&\seq& (\gamma_a)_\alpha{}^\beta \bar
\Upsilon_{\beta
i}\ \ ,\quad \{\bar Q_{\alpha i},\Upsilon_{\beta j}\}\seq -\half
(\gamma_c)_{\alpha\beta}
\epsilon^{abc}F_{ab}C_{ij}\  .\eea
As we will see next, it is important that this spin-$\half$ is charge
is a $SU(2)$ doublet.

Unlike the \sbf theories, \spbf theories exist on curved manifolds.
This is because under twisting one of the spinor supersymmetry charges
becomes as Lorentz scalar.  It is only this (identified as the BRST
charge) and any other scalar super-charges which must be maintained
by the twisting process.  Thus, all of the symmetry transformations
above, will not be needed in order to obtain the \sbf theory.  As the
twisting of D=3, N=4 supersymmetric theories has been performed before
\cite{BDL} we will simply state the results of the operation for this
simple case.  The twisted action is
\be
S^\ct\seq \int d^3\sigma\, Tr[\half \epsilon^{abc} (B_a F_{bc}\mi
\Sigma_{ab}\psi_c)]\ \ .\label{TwistS}\ee
Under twisting, the first term has not changed.  The $SU(2)$ fermions
have been drastically altered, however.  First, they were replaced by
bi-spinors  which were then written as a one-form ($\psi$), a two-form
($\Sigma$) and a pair of zero-forms.  Then two of the the local gauge
symmetry parameters were used to remove the zero-forms. The remaining
two local symmetries are realized, in eqn. (\ref{TwistS})  as
$\delta\Sigma=\xi \dual\psi$ and $\delta\psi=\xi'\Sigma$ for two
arbitrary local parameters $\xi$ and $\xi'$.

The action (\ref{TwistS}) was proposed by us \cite{BroLuc} as a
cohomological theory for ordinary $BF$ theories. In that work we
observed that this action can be written as the anti-commutator
of a Grassmann-odd, scalar charge with a certain expression.  Now
we see hot it is connected to (\ref{SN4BF}) and (\ref{SN4red}).

\vskip 0.5truein\goodbreak
\section{Twisting Via Hodge Decomposition}\label{Twist}
\setcounter{equation}{0}
\medskip
\forcepar
We would now like to see how to obtain the \sbf gauge theory from the
action $S^\ct$.  Such a construction requires a new addition to the
twisting procedure; namely, Hodge decomposition in a flat connection
background.  In this section, we will work on arbitrary closed,
orientable $3$-manifolds, $M$, with metric.  It is then convenient to
re-write $S^\ct$ in terms of forms as
\be
S^\ct\seq\int_M Tr(B\wedge F \mi \Sigma\wedge \psi)\ \
.\label{StwistForm}\ee
Furthermore, for convenience, we label the set of fields $\{B,A,\Sigma,
\psi\}$ as $\Xhat$ so that $S^\ct=S^\ct[\Xhat]$.  This action is invariant
\cite{BroLuc} under the set of symmetries
\bea
[\cq^H,B]&\seq&\psi\ \ ,\qquad \{\cq^H,\Sigma\}\seq F\ \ ,\nen
[Q^H,A]&\seq&\psi\ \ ,\qquad \{Q^H,\Sigma\}\seq \cd B\ \ ,\eea
where $\cd$ is the covariant exterior derivative.

\newpage
\subsection{Abelian Theory}
\medskip
\forcepar
To set the stage let us first focus attention on the abelian theory.
The only field which is a non-singlet under the $U(1)$ group is the
connection, $A$.  The Grassmann-odd, two-form, $\Sigma$, may be Hodge
decomposed as
\be
\Sigma ~\equiv~ d\chi \pls \dual d\eta \pls \sigharm \ \
,\label{HodgeAbel}\ee
where $\chi$, $\eta$ and $\sigharm $ are Grassmann-odd, $1-$, $0-$
and harmonic $2-$ forms, respectively, $d$ is the exterior derivative
on $M$ and $\dual$ is the Hodge dual operator.  An important point is
the absence of $\chi$ and $\eta$ zero-modes in this decomposition; we
will return to this point below.  Using this eqn. (\ref{HodgeAbel}) in
the action (\ref{StwistForm}), we find that its partition
function is\footnote{Wedge products are understood in the equations to follow.}
\be
\cz \seq \int[dB][dA][d\psi][d\chi][d\eta][d\sigharm] \cj
\exp{-\left\{\int_M \left[ BF - \chi
d\psi + \eta d\dual\psi + \sigharm \psi\right]\right\}}\
\ ,\label{ZY}\ee
where $\cj$ is the Jacobian for the change of variables from the
set $\Xhat$ to its Hodge decomposition, $\Yhat$, in which $\Sigma$
is replaced by the triplet $(\chi,\eta,\sigharm )$: $\Yhat=\{B,A,
\psi,\chi,\eta,\sigharm\}$. In the partition function for the
action $S^\ct[\Xhat]$, the functional integral over the $\Sigma$
and $\psi$ fields is equal to one. Thus we determine the Jacobian by
requiring that the functionalintegral over the Grassmann odd fields
in (\ref{ZY}) is also one. That is,
\be
\cz_{\Sigma\psi}\seq\int \left[d\chi,d\psi,d\eta,d\sigharm \right] \cj
\exp{\left\{\int_M \left[  \chi d\psi - \eta d\dual\psi -
\sigharm \psi\right]\right\}}\equiv 1\ \ .\label{ZSPa}\ee
At this stage we must be careful about the measures over these fermionic
fields.  As mentioned previously,  in the right-hand-side of equation
(\ref{HodgeAbel}), only those fields which are not in $\ker{d}$, appear.
In otherwords, there are $\chi$ and $\eta$ zero-modes present in $\Yhat$.
Yet, when we write the functional integral $\int [d\chi]$, etc., we allow
for $\chi$ to take values also in $\ker{d}$.  This means that we must
define the measures $[d\chi]$ and $[d\eta]$  so that they are equivalent
to functional integrals over the non-zero mode parts of the respective
fields.  We do this by noting that since these fermionic zero-modes are
absent from the action,  we can simply get rid of the functional integral
over them by inserting a complete set in the partition function.
Consequently, henceforth, by $[d\chi]$ and $[d\eta]$, we will mean that
the respective zero-modes have been inserted in the partition function.
Notice that we do not do the same for $\psi$ as its zero-modes explicitly
appear in the action in (\ref{ZY}).  Elaboration on this may be found
below. The action in the partition function is invariant under the $1$-form
symmetry. After gauge fixing the latter, we find that
\be
\cz_{\Sigma\psi}\seq b_{(1)}T  \det[\triangle_{(0)}]\ \ ,\ee
where $b_{(1)}=\dim(H^{(1)}(M))$ (if $dim H^{(1)}(M)=0$ then
$b_{(1)}=1$) and $T $ is the R-S torsion on the three dimensional
manifold, $M$. The factor of $b_{(1)}$ arises from the integral over
the harmonic $2$-form, $\sigharm $.  Being Grassmann-odd, it pulls
down that part of $\psi$ which is in $H^{(1)}(M)$.  These are
Grassmann-odd zero-modes of the exterior derivative on $1$-forms:
$\psi_{(0)}{}^I, ~I=1,\ldots, b_{(1)}$.   Our ansatz for $\cj$ is
then \be \cj= \inv{b_{(1)}} T^{-1}(3) {\rm det}^{-1}\left[
\triangle_{(0)}\right]\
.\ee
We notice that the last factor can be represented as the Gaussian
functional integral of two scalars, $\lambda$ and $\phi$.  Putting this
together with eqn. (\ref{ZY}) leads to
\bea
\cz&=&T^{-1}(3)\int [dA,dB,d\psi,d\chi,d\eta,d\lambda,
d\phi,d\eta',d\lambda',d\phi' ] \prod_{I=1}^{b_{(1)}} \psi_{(0)}{}^I
e^{-S_{SBF}}\ ,\nen
S_{SBF}&=& \int_M\left[BF\mi \chi d\psi\pls \eta d\dual\psi \pls
\lambda\triangle\phi \pls \eta' d\dual\chi\pls \lambda'\triangle \phi'\right]\
\ ,\eea
where the pair $\lambda'$ and $\phi'$ are the anti-ghost and ghost for
the gauge-fixing of the $1$-form symmetry on $\chi$.  $S_{SBF}$ is the
action for the first-order (super-$BF$) form of the Donaldson-Witten
theory of flat $U(1)$ connections on three manifolds.  The global
symmetries mentioned above are manifested in this action as
\bea
\trans{\cq^H}{B}{\psi}\ \ ,\qquad \trans{\cq^H}{\chi}{ -A}\ \ ,\nen
\trans{Q^H}{A}{\psi}\ \ ,\qquad \trans{Q^H}{\chi}{-B}
\ \ .\label{GlobalSym}
\eea
As before, they are nilpotent.
\bigskip\goodbreak
\subsection{Non-Abelian Theory}
\medskip
\forcepar
The preceding discussion carries over to the non-abelian theory with
a few twists.  One will be  the appearance of potential (Yukawa-like
coupling) terms in the action.   The other has to do with the Hodge
decomposition of $ad(G)$ valued forms on $M$.  We address the latter
first.  Writing $\Sigma$ as
\be
\Sigma ~\equiv~ \cd\chi \pls \dual \cd\eta \pls \sigharm \ \
,\label{NAHODGE}\ee
is not an orthogonal decomposition, in general.  However, it satisfies
the latter criteria if the connection is flat.  Since our partition function
has support only on flat connections,  (\ref{NAHODGE})  is valid in this
context.  Equations (\ref{ZY}-\ref{ZSPa}) also hold here except that the
exterior derivative is replaced by $\cd$ and there are traces over the
bracketed terms in the action.  Having decomposed $\Sigma$, we must now
determine the transformations on the new fields.  In order to do this,
it is best to first include the Yang-Mills transformation laws in the
action of one of the fermionic generators.  By doing this, we identify
that the action is invariant under the transformations generated by
$\cq$ and $Q\equiv Q^H+Q_{YM}$:
\bea
&&\trans{\cq}{B}{\psi}\ \ ,\qquad ~
\trans{\cq}{\Sigma}{F}\ \ ,\nen
\vbox{\vskip 0.6truecm}
&&\trans{Q}{A}{\psi-\cd\Theta}\ \ ,\qquad \quad
\trans{Q}{\Sigma}{\cd B+[\Theta,\Sigma]}\ \ ,\nen
&&\trans{Q}{B}{[\Theta,B]}\ \ ,\qquad\qquad ~
\trans{Q}{\psi}{[\Theta,\psi]}\ \ .\label{QTOT}
\eea
The commutators on the right-hand-sides of these expressions are those of
the gauge algebra. {}From these we read off that the action of the charges
on the new fields are
\bea
&&\trans{\cq}{\sigharm }{ - F}\ \ ,\nen
\vbox{\vskip 0.6truecm}
&&\trans{Q}{\chi}{[\Theta,\chi]-B}\ \ , ~
\trans{Q}{\eta}{[\Theta,\eta]}\ \ ,\nen
&&\trans{Q}{\sigharm }{[\Theta,\sigharm ] -
[\psi,\chi]+[\dual\psi,\eta]}\ \ ,\label{NEWQTOT}
\eea
and the ``horizontal'' charges act as
\be
\trans{Q^H}{A}{\psi}\ \ ,~
\trans{Q^H}{\chi}{-B}\ \ , ~
\trans{Q^H}{\sigharm }{[\dual\psi,\eta]-[\psi,\chi]}\ \ .\label{NEWQ}
\ee
As in the abelian case, there is a local symmetry manifest in the Hodge
decomposition under which we can shift $\chi$ by a covariantly exact
quantity.  We will fix this symmetry later.

The action  obtained by combining (\ref{NAHODGE}) and (\ref{StwistForm}) is
\be
S^\ct[\Yhat]\seq \int_M Tr\left(BF\mi \chi \cd \psi + \eta \cd \dual\psi +
\sigharm \psi\right)\ \ .\label{SNAHD}\ee
If we integrate over the harmonic form $\sigharm $, we find two things.
First, the effect of performing the Grassmann-odd integral is to pull down
from the action, and into the measure, the zero-modes of $\psi$ which are
elements of $H^1(M,G)$.  Given the presence of fermionic zero-modes, we
might expect the symmetries to be anomalous.  This is indeed the case,
as the remaining action given by all but the last term in (\ref{SNAHD})
is not invariant under the transformations (\ref{NEWQ}).  Our situation
is reminiscent of deriving a component level supersymmetric action by
bootstrapping our way term by term.  A simple counting of off-shell
degrees of freedom shows that we have one more $\psi$ degree of freedom
than we have in $A$; a similar discrepancy holds between the $(\chi,\eta)$
fields and $B$.  Thus in order to match the number of fermionic and
bosonic degrees of freedom, we must add a boson to each of the $(B,\chi,
\eta)$ and $(A,\psi)$ supermultiplets.  Let us call these $\lambda$
and $\phi$, respectively,  so that we have the new supermultiplets
\footnote{The $A$ supermultiplet will eventually be enlarged to include
an additional Grassmann-odd scalar.} $(A,\psi,\phi)$ and $(B,\chi,\eta,
\lambda)$.  Then, in order to make the action invariant under (\ref{NEWQ})
we must add to it two terms, $\lambda \cd \dual \cd \phi$ and $\lambda [\psi,
\dual\psi]$, and enlarge the symmetry transformations to include
$\trans{Q^H}{\lambda}{-\eta}$ and $\trans{Q^H}{\eta}{[\phi,\lambda]}$.
With these new terms in the action, we have deduced part of the Jacobian.
So we write $\cj=\cj'\int[d\lambda][d\phi] \exp{\{-\int_M \lambda \cd
\dual \cd \phi+ \lambda [\psi,\dual\psi]\}}$.  It now remains for us to
determine $\cj'$.  Although we had in mind that the partition function
has support only on flat connections when expanding $\Sigma$, we can
consider the connection in the action to be arbitrary.  In this case, the
latter is invariant under the local symmetry $\delta \chi = -\cd \Lambda$,
$\delta B =[\Lambda,\psi]$.  Clearly, this is a symmetry of the action
if $A$ is a flat connection also. In any case, it must be fixed which we
do by selecting the gauge slice $\cd \dual\chi=0$.  This leads to the action
\bea
S_{SBF}\seq \int_M Tr\bigg(&&\hskip -1truecm BF\mi  \chi
\cd \psi
\nen
&+&\eta \cd \dual\psi+\lambda \cd \dual \cd \phi+ \lambda [\psi,\dual\psi]
\nen
&+& \eta' \cd \dual\chi + \lambda'\cd \dual \cd \phi'+\lambda'
[\psi,\dual\chi]\,\bigg)\ \ ,\eea
where (as in the abelian case) $\eta'$ is the gauge-fixing Lagrange
multiplier and $(\lambda')\phi'$ is the  corresponding (anti-)
ghost\footnote{The supermultiplets are now $(A,\psi,\eta',\phi,\lambda')$
and $(B,\chi,\eta,\lambda,\phi')$, with equal numbers of fermionic and
bosonic degrees of freedom.}.  It is important to note that we are
justified in replacing the flat connection in the covariant derivative by
$A$ as the path integral has support only on flat connections.  We can
determine the remaining factor in the Jacobian, $\cj'$.  Recall that we
must impose
\bea
\cz_{\Sigma\psi}&=& \int [d\Sigma][d\psi] e^{-\int_M
Tr(\Sigma\wedge\psi)}\nen
&=& b^{(1)}(G) \int [d\chi][d\eta] [d\psi] [d\eta']
[d\lambda][d\phi][d\lambda'][d\phi'] e^{-S_{SBF}} \cj\ \ .\eea
The bottom line of this expression may be evaluated if we assume that
there are no non-trivial zero-modes of the  scalar laplacian, $\Delta_A
{}^{(0)}$.  The integrals over $\phi$ and $\phi'$ lead to $\delta (
\Delta_A\lambda)$ and $\delta (\Delta_A\lambda')$ respectively.  With
the no zero-mode assumption, these set $\lambda$ and $\lambda'$ to be
zero, thereby removing the Yukawa couplings.  The rest of the integrals
are then Gaussians.  After diagonalizing them, we find $\cz_{\Sigma
\psi}= b^{(1)}(G) \cj T(A)$ where $T(A)$ is the R-S torsion\footnote{Strictly
speaking, the R-S torsion is obtained from these integrals only after the
$B$ integral is performed as we need the flat connection condition in order
to obtain this topological invariant.} and $b_{(1)}(G)=\dim(H^1(M,G))$.
Hence we have finally found the sought after new representation of the
twisted SUSY-$BF$ partition function:
\bea
{\cal Z}=\inv{b^{(1)}(G)}\int&&\hskip -0.7truecm [dA][dB]
[d\chi][d\eta]
[d\psi] [d\eta'] [d\lambda][d\phi][d\lambda'][d\phi']
\prod_{I=1}^{b^{(1)}(G)}
\psi_{(0)}^I\times\nen
&&\times ~T^{-1}(A) e^{-S_{SBF}}\ .\eea
We recognize this as the partition function for the super-$BF$ theory with
the ratio of determinants which appears in the inverse R-S torsion (but with
the flat connections replaced by the general connection, $A$) inserted.  In
addition, the $\psi$ zero-modes appear naturally inserted.  The form-degrees,
Grassmann-parity and corresponding ghost numbers of the fields which we have
introduced in order to write this partition function are given Table I.
\begin{center}
\begin{tabular}{|c|c|c|c|}\hline
FIELD&DEGREE&G-PARITY&GHOST \#\\ \hline\hline
$B$&$1$&even&~~0\\ \hline
$A$&$1$&even&~~0\\ \hline
$\chi$&$2$&odd&$-1$\\ \hline
$\psi$&$1$&odd&$~~1$\\ \hline
$\eta$&$0$&odd&$-1$\\ \hline
$\eta'$&$0$&odd&$~~1$\\ \hline
$\lambda$&$0$&even&$-2$\\ \hline
$\phi$&$0$&even&$~~2$\\ \hline
$\lambda'$&$0$&even&$~~0$\\ \hline
$\phi'$&$0$&even&$~~0$\\ \hline
\end{tabular}
\end{center}
Collecting the various pieces of the $Q^H$
transformation we find
\bea
&&\trans{Q^H}{A}{\psi}\ \ ,\qquad \trans{Q^H}{\chi}{B+\cd \phi'}\ \
,\qquad
\trans{Q^H}{\psi}{\cd \phi}\ \ ,\nen
&&\trans{Q^H}{\lambda}{-\eta}\ \ ,\qquad
\trans{Q^H}{\eta}{[\phi,\lambda]}\ \
,\nen
&&\trans{Q^H}{\lambda'}{-\eta'}\ \ ,\qquad
\trans{Q^H}{\eta'}{[\phi,\lambda']}\
\ .\eea
{}From this we read-off that  the square of $Q^H$ is a gauge transformation.
The action, $S_{SBF}$,  is known to be $Q^H$ -exact.
\vskip 0.5truein
\section{Concluding Remarks}\label{Conc}
\medskip
\forcepar
In this paper, we have constructed the D=3, N=4 super Yang-Mills superspace
geometry and used it to construct the corresponding off-shell SUSY-$BF$
gauge theory.  We have found that, generically, SUSY-$BF$ gauge theories
do not have dynamical fermions; yet, super-BF theories do.  In order to
twist the 3D, N=4 SUSY-$BF$ theory we found it necessary to Hodge decompose
one of the twisted fermions.  Additionally, we have seen that in order for
the fermion and boson determinants to cancel (up to signs) in the partition
function of the SUSY-$BF$ theories, we should re-defined the measure to
include the Ray-Singer analytic torsion but with flat connections replaced
by the full (quantum) connection. After twisting and then Hodge decomposing
the SUSY-$BF$ action, we found that the partition function of the SUSY-$BF$
theory becomes that of the super-BF theory but with the same ratio of
determinants of covariant laplacians, which appears in the inverse Ray-Singer
torsion, inserted in the measure.  It would appear from these results that
the only difference between $BF$ and super-BF gauge theories is a choice
of vacuum.
\vskip 0.5truein
\appendix
\centerline{\Large\bf \underline{Appendix}}
\section{Off-shell 3D, N = 4  Supergravity}
\medskip
\forcepar
The  construction of 3D, N = 4 supergravity is essentially equivalent to
the problem of finding the consistent truncation of 4D, N = 2 supergravity
to three dimensions.  The fact that this latter problem is a long solved
one \cite{deWit} provides a quick and handy technique for resolution of the
three dimensional one.  Let us go through the logic that leads to our result.
In the off-shell 4D, N = 2 supergravity theory, the supermultiplet can be
viewed as the direct sum of two 4D, N = 1 supermultiplets.  One of these
supermultiplets is the irreducible non-minimal off-shell 4D, N = 1
supergravity supermultiplet \cite{Breit}.  The other multiplet is the
off-shell 4D, N = 1 matter gravitino supermultiplet \cite{Siegat}.  Both of
these supermultiplets contain 20-20 bosonic and fermionic degrees of freedom.
Of these degrees of freedom, there are 4-4 propagating physical degrees of
freedom.  In 3D, both supergavity and matter gravitino supermultiplets must
consist solely of auxiliary degress of freedom.  This tells us that the
truncation to 3D must be such that it separates all of the physical degrees
of freedom from the 3D, N = 2 supergravity and matter gravitino
supermultiplets.
The physical degrees of freedom in the case of each supermultiplet wind up
in separate 3D, N = 2  vector supermultiplets.   Thus, we conclude that the 3D,
N = 2 non-minimal supergravity supermultiplet consist of 16-16 bosonic and
fermionic degrees  of freedom.   The same holds true for the 3D, N = 2  matter
gravitino multiplet.  Now we simply argue that the direct sum of the 3D, N = 2
supergravity and matter gravitino supermultiplets must correspond to the 3D,
N = 4 supergravity multiplet.  This observation by itself totally determines
the spectrum of the theory we want to construct for the 3D, N = 4 case.  It
must
consist of 32-32 bosonic and fermionic degrees of freedom.   Since we have
argued
that this theory can be directly obtained by the dimensional reduction of the
4D,
N = 2 supergravity theory along with the truncation described in the paragraph
above, we even have {\it {a priori}} knowledge of the spectrum of the theory.
The
knowledge of the spectrum is not sufficient. We also need to know the
transformation laws of the supermultiplet. This will be subject of a separate
report.

The implementation of this construction at the level of actions
is simple. We start with the 4D, N = 4 supergravity action which we
represent as
$$ \int d^4 x {\cal L}_{N = 2, {\rm SG}} = \int d^4 x [~ {\cal L}_{phys}
{}~+~
{\cal L}_{aux} ~]  \eqno(A.1) $$
where ${\cal L}_{phys} $ contains the physically propagating degrees of
freedom $e_{a}{}^m$, $\psi_{a} {}^{\a i}$,  ${\bar \psi}_{ a} {}^{\dot \a}
{}_{ i}$,  $B_a $ and ${\cal L}_{aux} $
contains all of the auxiliary fields. Under a toroidal compactification,
these fields ``split'' according to the following table.
\begin{center}
\renewcommand\arraystretch{1.2}
\begin{tabular}{|c|c| }\hline
 $4{\rm D}$ $ $  & $~~~$ $3{\rm D}$ $  $
  \\ \hline
 ${\rm SG} $ ${\rm field }$  & ${\rm SG}$ ${\rm fields }$ $ ~~|~~ $
${\rm Matter}$ ${\rm fields }$
  \\ \hline\hline
 $   e_a {}^m  $ &   $ {~~~~~~~~~~~} e_a {}^m ~~~~~|~~~   b_a  \equiv
e_a {}^3
 ,~ \phi
\equiv  e_3 {}^3$  \\ \hline
$  \psi_{a} {}^{\a i} $ &   $ { ~~~~~~~~~} \psi_{a} {}^{\a i} ~~~~|~~~~~~
\varphi
{}^{\a i} \equiv  \psi_{3} {}^{\a i} ~~~ $  \\ \hline
 $  B_a   $ &   $ {~~~~~~~~~~~} ~~|~~~   B_a  ,~ b \equiv  B_3   $  \\ \hline
\end{tabular}
\end{center}
\vskip.2in
\centerline{{\bf Table II}}
\noindent
As explcitly seen, two 3D, N = 2 vector multiplets appear as matter fields.
This implies that in $ {\cal L}_{aux}$, two scalar auxiliary fields are
associated with the propagating matter fields in our table. These fields
are thus part of the two 3D, N = 2 vector multiplets. After truncating out
these two vector multiplets we are left with an action
$$ \int d^3 \sigma {\cal L}_{N = 4, {\rm SG}} \eqno(A.2)
$$
The explicit presentation of these results, as well as generalizations
involving 3D, N = 4 CS supergravity theory, will be given elsewhere.

\newpage
\section {3D, N = 4  Superconformal Algebra }
\medskip
\forcepar
Below we give the form of the superconformal algebra that is associated with
the three dimensional off-shell supergravity theory discussed in our text.
For the sake of simplicity, only the non-vanishing graded commutators are
listed below.

$$
[ {\cal M}_{a } , {\cal M}_{b } \}  =   \epsilon_{a b}{}^{c } {\cal M}_{c }
{}~~,
{}~~
[ {\cal M}_{a } , { P}_{b } \}  =   \epsilon_{a b}{}^{c } { P}_{c } ~~, ~~
[ {\cal M}_{a } , { K}_{b } \}  =   \epsilon_{a b}{}^{c } { K}_{c } ~~,
\eqno(B.1)$$

$$
[  {\cal D} , {P}_{a }  \}  =   {P}_{a } ~~,~~
[  {\cal D} , {K}_{a }  \}  = - {K}_{a } ~~,~~
\eqno(B.2) $$

$$
[ {P}_{a } , {K}_{b } \}  ~=~  \frac 12 \eta_{a b}  {\cal D}
 ~-~ \frac 12 \epsilon_{a b}{}^{c } {\cal M}_{c } ~~, ~~
\eqno(B.3) $$

$$
[ {Q}_{\a i} , {\bar S}_{\b }{}^j \}  = C_{\a \b} [ ~\d_i {}^j ( {\cal D} +i
{\cal Y} ) ~-~ {\cal T}_i {}^j ~] ~-~ i \d_i {}^j (\g^c)_{\a \b} {\cal M}_{c }
\eqno(B.4) $$

$$
[ {\cal M}_{a } , {Q}_{\a i}  \}  = i \frac12  (\g_a)_{\a }{}^{\b} {Q}_{\b i}
{}~~, ~~
[ {\cal M}_{a } , {S}_{\a i}  \}  = i \frac12  (\g_a)_{\a }{}^{\b} {S}_{\b i}
{}~~,
\eqno(B.5) $$

$$
[ {P}_{a } , {S}_{\a i}  \}  =  \frac12  (\g_a)_{\a }{}^{\b} {Q}_{\b i} ~~, ~~
[ {K}_{a } , {Q}_{\a i}  \}  =  \frac12  (\g_a)_{\a }{}^{\b} {S}_{\b i} ~~,
\eqno(B.6) $$

$$
[ {Q}_{\a i} , {\bar Q}_{\b }{}^j \} = 2 \d_i {}^j  (\g^c)_{\a \b} {P}_{c }
{}~~,~~
[ {S}_{\a i} , {\bar S}_{\b }{}^j \} = 2 \d_i {}^j  (\g^c)_{\a \b} {K}_{c }
{}~~,~~
\eqno(B.7) $$

$$
[  {\cal Y} , {Q}_{\a i}  \}  = i  \frac12  {Q}_{\a i} ~~,~~
[  {\cal Y} , {\bar Q}_{\a}{}^{ i}  \}  =  - i  \frac12
{\bar Q}_{\a}{}^{  i} ~~,~~
\eqno(B.8) $$

$$
[  {\cal Y} , {S}_{\a i}  \}  = -  i\frac12  {S}_{\a i} ~~,~~
[  {\cal Y} , {\bar S}_{\a}{}^{ i}  \}  =   i  \frac12
{\bar S}_{\a}{}^{  i} ~~,~~
\eqno(B.9) $$

$$
[  {\cal D} , {Q}_{\a i}  \}  =  \frac12  {Q}_{\a i} ~~,~~
[  {\cal D} , {S}_{\a i}  \}  = - \frac12  {S}_{\a i} ~~,~~
\eqno(B.10) $$

$$
[  {\cal T}_i {}^j , {Q}_{\a k}  \}  = \d_k {}^j {Q}_{\a i} ~-~ \frac 12
\d_i {}^j {Q}_{\a k}  ~~,~~
[  {\cal T}_i {}^j  , {S}_{\a k}  \}  = \d_k {}^j {S}_{\a i} ~-~ \frac 12
\d_i {}^j {S}_{\a k}  ~~,~~
\eqno(B.11) $$

$$
[  {\cal T}_i {}^j  ,  {\cal T}_k {}^l  \}  = \d_k {}^j {\cal T}_i {}^l
{}~-~  \d_i {}^l {\cal T}_k {}^j ~~.
\eqno(B.12) $$

\vskip 0.5truein
\section{The 3D, N = 4 Supersymmetric U(1) Anyonic Model}
\medskip
\forcepar
One of the nice feature of possessing off-shell supersymmetric
representations is that they can be added freely to other such models
without regard to loss of supersymmetry. In particular, our construction
of the 3D, N = 4 sBF (supersymmetric BF) action is such that it can easily
be coupled to 3D, N = 4 scalar multiplets without new difficulties
arising. Thus, we are able to extend the constuction of supersymmetric
anyonic models to the level of N = 4 supersymmetry.  Our construction
below marks the first time this has been achieved.

To succeed in this effort requires an off-shell 3D, N = 4 scalar
supermultiplet. Fortunately, such a representation is available after
a little bit of thought.  The secret to finding this representation
is to recall that one off-shell 4D, N = 2 scalar supermultiplet
\footnote{Actually there is the so-called harmonic space formulation
of this supermultiplet \cite{Ivan}. But this leads to a model with an
infinite set of auxiliary fields.} is known in the physics literature,
the relaxed hypermultiplet \cite{HoST}. Using the technique of toroidal
compactification leads to a 3D, N = 4 scalar supermultiplet!  Furthermore,
the first of our two vector (since it may be regarded as the dimensional
reduction of the 4D, N = 2 vector supermultiplet) supermultiplets in (3.1)
may be freely coupled to the relaxed hypermultiplet. This opens the way
for us to couple our supersymmetric BF action to matter and forming a N = 4
anyonic type of model.

The superspace action for our N = 4 anyonic theory follows immediately
from the corresponding 4D, N = 2 theory. The total action consists
of the super BF action in section in (3.7) added to the following
superfield action\footnote{We have adhered to the conventions of
\cite{HoST} in the names of quantities below. We warn the reader
in particular that the symbols $\lambda$ and $\rho$ denote superfields
below and are not related to the component field given the same
names in the body of this work.}
$$ \int d^3 x d^8 \theta ~[ ~ (\lambda_{\a}{}^i \rho^{\a} {}_i ~+~
\psi_{\a}{}^i \s^{\a} {}_i ~+~ L^{ijkl} X_{ijkl} )~~+~~ {\rm c.}{\rm c.}
{}~] \eqno(C.1) $$
In this expression the fundamentally unconstrained superfield
potentials are $ \rho^{\a} {}_i $, ${\bar \s}_{\a} {}^i $ and
$X_{ijkl}$.  All other superfields associated with the 3D, N = 4
relaxed hypermultiplet are expressed in terms of these fundamental
fields as
\newline
$ {~~} L^{i j} ~=~ {\bar \nabla}^{i j} {\nabla}^{3 k}{}_{\a}  \rho^{\a} {}_k
{}~-~ {\nabla}^{i j} {\bar \nabla}^{3}{}_{k \a}  {\bar \s}^{\a k} $
\newline
$ {~~~~~~~~~~~}~-~ {\nabla}^{i j} {\bar \nabla}^{3}{}_{k \a} \rho^{\a}
{}_k
{}~-~{\nabla}_k {}^{ ( i | } {\bar \nabla}^{j) k} {\bar \s}^{\a k}
{\bar \nabla}_{\a l}{\bar \s}^{\a l} $
$$~~~-~ i \frac 12 {\bar \nabla}^{i j} ( {\bar W} {\nabla}^{ k}{}_{\a}
\rho^{\a} {}_k ) ~-~ i \frac 12 {\nabla}^{i j} ( W
{\bar \nabla}_{ k}{}^{\a} {\bar \s}_{\a} {}^k  )  ~~~, {~~~~~~~~~~~~~}
{~~~~~~~~~~}
\eqno(C.2)
$$

$$ \lambda_{\a} {}^i \equiv {\nabla}_{\a j} L^{i j} ~~,~~
{\bar \psi}_{\a} {}^i \equiv {\nabla}_{\a j} L^{i j} ~~.~~
\eqno(C.3)
$$

$$ L^{ijkl} ~=~ - \frac 25 {\nabla}^{ ( i  j| }  {\bar \nabla}^{ |k l ) }
(~ {\nabla}^{m}{}_{\a} \rho^{\a} {}_m ~+~ {\bar \nabla}_{m}{}^{\a}
\s_{\a} {}^m  ~) ~~~. {~~~~~~~~~~} {~~~~~~~~~~}
\eqno(C.4)
$$
In these expressions $ {\nabla}_{\a i}$ is the gauge covariant derivative that
appears in (3.1) and ${\bar W}$ is the corresponding field strength.

\end{document}